\documentclass[12pt,amsfonts]{article}

\usepackage{amsmath}

\usepackage{amssymb}
\usepackage{bm}
\usepackage{graphicx}
\usepackage{xcolor}
\usepackage{caption}
\usepackage{subcaption}
\usepackage{appendix}
\usepackage{bigints}
\usepackage{yfonts}
\usepackage{bbold}
\usepackage[T1]{fontenc} 
\usepackage{amssymb}
\usepackage{bm}
\usepackage{graphicx}
\usepackage{xcolor}
\usepackage{caption}
\usepackage{subcaption}
\usepackage{appendix}
\usepackage{bigints}
\usepackage{yfonts}
\usepackage{bbold}

\newcommand{\be}{\begin{equation}}
\newcommand{\ee}{\end{equation}}
\newcommand{\ba}{\begin{array}}
\newcommand{\ea}{\end{array}}
\newcommand{\bqa}{\begin{eqnarray}}
\newcommand{\eqa}{\end{eqnarray}}
\newcommand{\bea}{\begin{eqnarray}}
\newcommand{\eea}{\end{eqnarray}}

\begin{document}

\newtheorem{defi}{Definition}[section]
\newtheorem{lem}[defi]{Lemma}
\newtheorem{prop}[defi]{Proposition}
\newtheorem{theo}[defi]{Theorem}
\newtheorem{rem}[defi]{Remark}
\newtheorem{cor}[defi]{Corollary}

\title{On Quasi-integrable Deformation Scheme of The KdV System}
\author {Kumar Abhinav$^1$\footnote{E-mail: {\tt kumar.abh@mahidol.ac.th}} and Partha Guha$^2$\footnote{E-mail: {\tt partha.guha@ku.ac.ae}}\\$^1$Centre for Theoretical Physics \& Natural Philosophy\\
``Nakhonsawan Studiorum for Advanced Studies", Mahidol University,\\
Nakhonsawan Campus, Phayuha Khiri, Nakhonsawan 60130, Thailand\\$^2$Khalifa University of Science and Technology,\\
PO Box 127788, Abu Dhabi, UAE}
\date{\today}
\maketitle
\abstract{We put forward a general approach to quasi-deform the KdV equation by deforming the corresponding Hamiltonian. Following the standard Abelianization process based on the inherent $sl(2)$ loop algebra, an infinite number of anomalous conservation laws are obtained, which yield conserved charges if the deformed solution has definite space-time parity. Judicious choice of the deformed Hamiltonian leads to an integrable system with scaled parameters as well as to a hierarchy of deformed systems, some of which possibly being quasi-integrable. As a particular case, one such deformed KdV system maps to the known quasi-NLS soliton in the already known weak-coupling limit, whereas a generic scaling of the KdV amplitude $u\to u^{1+\epsilon}$ also goes to possible quasi-integrability under an order-by-order expansion. Following a generic parity analysis of the deformed system, these deformed KdV solutions need to be parity-even for quasi-conservation which may be the case here following our analytical approach. From the established quasi-integrability of RLW and mRLW systems [Nucl. Phys. B {\bf 939} (2019) 49–94], which are particular cases of the present approach, exact solitons of the quasi-KdV system could be obtained numerically.}


\bigskip
\noindent{\bf Keywords and Keyphrases}: Quasi-integrable deformation, KdV equation, NLS equation.

\bigskip

\section{Introduction}
The $1+1$-dimensional Korteweg-de Vries (KdV) equation \cite{NRE1} is applicable to many real-life phenomena such as flow of a shallow fluid. Being third order in space derivative, though this non-linear partial differential equation (PDE) has a dispersion odd in momentum powers, it is completely integrable \cite{2} and supports localized soliton solutions \cite{NRE2} that can represent different observable physical objects in fluid dynamics like tidal waves. Solitonic structures are well-known in other nonlinear systems in $1+1$-dimensions such as the nonlinear Schr\"odinger (NLS) equation which is also integrable and may seem to be more closely related to physical phenomena as the NLS solitons have been observed in Bose-Einstein condensates, cold atoms and optics \cite{NRE3}. The NLS with quadratic dispersion is more suitable for a `physical visualization' is fundamentally different from the KdV system. The KdV is geometrically connected to diffeomorphism group \cite{Arnold} whereas NLS is tagged with loop algebra \cite{Chau}. Moreover the {\it usual} Lax representation \cite{Lax} of KdV system involves second and third order monic differential operators ($L,A$), whereas that of the NLS system is given by $2\times 2$ matrices \cite{2}. However, it is known that in a  suitable weak-coupling limit, their respective solutions map into each-other \cite{5,Jun}, including their soliton solutions. 
\paragraph*{}The NLS soliton dynamics has been well-studied including its various deformations \cite{NRE3} in various physical systems and same can be said about the exact KdV system \cite{NRE4}. However, detailed studies of localized structures for deformations of the KdV system are sparse which could be due to it being third-order in space-derivative that effects its solvability more severely. Deformations of such continuous systems may not be integrable in general as the delicate balance between dispersion and nonlinearity is crucial for the infinitely many conserved quantities (charges) hallmarking their integrability. Due to such high sensitivity, most of these deformations do not posses any conserved charges. Subsequently, it becomes very difficult to obtain localized solutions for these deformed systems since the integrable solitons derive their robustness of existence from infinitely many conservation laws \cite{NRE5}.
\paragraph*{}On the other hand, real physical systems are characterized by finite number of degrees of freedom, prohibiting integrability of the corresponding field-theoretical models in principle. Yet, they are physically known to posses solitonic states, very similar in structure to the integrable ones. Some examples include particular deformations of sine-Gordon (SG) \cite{FerrSG}. This motivates the study of continuous systems as slightly deformed integrable models. In a recent work \cite{FZ,FZ1}, the SG model was shown to be deformable into an approximate system that supports the conservation of only a subset of the charges while the others behaving anomalously, corresponding to an anomalous zero-curvature condition. Such behavior was seen in the supersymmetric extension to the SG system \cite{N1} and for certain deformations of the NLS \cite{FZ2} and AB \cite{Own3} systems also. Moreover, the anomalous charges are seen to regain conservation for localized solutions which are far apart. In particular cases single and multi-solitonic solutions were numerically obtained for these deformations \cite{FerrSG,FZ,FZ1,FZ2}. Expectantly the corresponding charges were anomalous when these solitonic structures interacted locally, but when the latter are well-separated these charges return to being conserved. This was interpreted as {\it asymptotic integrability} and the corresponding systems are deemed as quasi-integrable (QI). 
\paragraph*{}For the particular case of NLS equation, Ferreira {\it et. al.} \cite{FZ2} modified NLS {\it potential}, the term in the Hamiltonian that leads to the nonlinearity in the NLS equation as $V(\varphi)=\vert\varphi\vert^2\to\left(\vert\varphi\vert^2\right)^{2 + \epsilon}$. Here $\varphi$ satisfies the NLS equation and $\epsilon$ is the deformation parameter. It was found that this model possesses an infinite number of quasi-conserved (anomalous) charges that are conserved asymptotically corresponding to numerically obtained solitonic structures. It was further found that the anomaly functions corresponding to the deformed curvature and that for the anomalous charge evolution have definite parity properties essential for the asymptotic integrability of the system. Owing to the closeness with the NLS system \cite{5,Jun}, the KdV system is also expected to display such quasi-integrability. However this third order equation in space eludes a direct dynamical interpretation in the usual classical sense and there are no `potential' analogue here. More importantly the Lax formalism is essential to the quasi-integrability mechanism wherein the inherent $SU(2)$ structure leading to an $sl(2)$ loop algebra is utilized \cite{FerrSG,FZ,FZ1,FZ2}. The usual KdV Lax pair is made of monic differential operators devoid of such algebra. $SU(2)$ representations of the KdV Lax pair, however, exists \cite{1} with proper grade structure \cite{NN001} for the Abelianization procedure \cite{FerrSG,FZ} required for quasi-integrability. Subsequently, a general structure for the quasi-integrable deformation of the KdV system is yet to be obtained. 
\paragraph*{}In a recent and important work \cite{KdVQI1} particular deformation of the KdV equations, which can be identified with non-integrable systems such as the regularized long-wave (RLW) \cite{RLW1,RLW2} and the modified regularized long-wave (mRLW) \cite{mRLW} equations, were shown to be quasi-integrable. Detailed analytical results, supported by numerical evaluation of single and multi-soliton structures, ensured asymptotic integrability of such systems for certain ranges of the deformation parameters. However, a general way to quasi-deform the original KdV system,, in the likeness of sine-Gordon \cite{FZ,FZ1} or NLS \cite{FZ2,AGM}, has not been proposed yet. In the present work we propose a general framework of deformation of the KdV equation that leads to quasi-integrability. We obtain the $sl(2)$ loop algebraic Abelianization \cite{FZ1,FZ2} of the KdV system and obtain the anomalous charges. We further provide the generic parity analysis of this deformation based on the said loop algebra to obtain definite parity structure of the quasi-deformation anomalies, crucial for the known cases. Finally a few of the deformed solutions and corresponding anomalies are analyzed approximately that conforms to the quasi-integrability structure.
\paragraph*{}As mentioned before, being a third order differential system, the KdV equation does not accommodate a {\it dynamical} deformation of the Lax pair. Instead, a more general {\it off-shell} deformation scheme, that of the KdV Hamiltonian has been achieved, allowing for suitably deformed Lax component, that further allows for a hierarchy of higher-derivative extensions of KdV. At the simplest level, such deformations lead to scaling of the KdV parameters and thus retaining integrability.
In the perturbative domain, as the KdV and NLS systems are related through a weak-coupling map \cite{5,Jun} between the solutions, we obtain a map between this quasi-KdV and the known quasi-NLS results \cite{FZ2}. We further infer about the connection of the quasi-KdV system to its non-holonomic (NH) deformation \cite{1}, the latter retaining integrability. Since the connection between quasi- and non-holonomic NLS system had been compared \cite{AGM} and are found to mutually correspond asymptotically, we expect a similar property for the respective deformations of the KdV system.
\paragraph*{}In the following, section \ref{Sec2} provides a detailed loop algebraic structure of the general quasi-deformed KdV system. In section \ref{Sec3} we obtain a general analysis of the deformation algebraic structure of the deformed system  ensuring quasi-conservation of the charges for localized solutions. We obtain some detailed results in the perturbative limit in section \ref{Sec4} with particular examples. Finally we discuss and conclude in section \ref{Sec5} highlighting remaining issues and further possibilities.

\section{Quasi-Integrable Deformation of KdV equation}\label{Sec2}

\subsection{Zakharov-Sabat representation}
It is well known that a systematic procedure of obtaining most finite dimensional completely integrable systems is Adler, Kostant and Symes ( AKS) theorem \cite{AvM,Sy} applying to some Lie algebra $\mathfrak{g}$ equipped with an ad-invariant non-degenerate bi-linear form. When this scheme is applied to loop algebra and the Fordy-Kulish decomposition scheme is invoked then the NLS \cite{FK} and KdV \cite{P1} equations can be formulated from there. This mechanism can also be applied for the construction of hierarchies too \cite{GM}. In case of the KdV system, the most general construction that can be derived from this AKS procedure is a pair of coupled complex KdV equations \cite{1}, through construction of the Lax pair:

\be
A=Q \quad{\rm and}\quad B=T+[S,Q],\label{1}
\ee
where,

\bea
&&T=-Q_{xx}+\left[Q^+,\left[Q^-,Q^+\right]\right]-\left[Q^-,\left[Q^-,Q^+\right]\right] \quad{\rm and}\nonumber\\
&&S=Q^+_x+Q^-_x+4c\left(Q^++Q^-\right),\qquad c\in{\mathbb R}.\label{2}
\eea
With the definitions,

\bea
&&Q=\left( \ba{cc} 0 & \bar{q} \\ -q & 0 \ea \right);\quad Q^+=\left( \ba{cc} 0 & \bar{q} \\ 0 & 0 \ea \right),\quad Q^-=\left( \ba{cc} 0 & 0 \\ -q & 0 \ea \right),\label{3N}\\
&&Q=Q^++Q^-\equiv \bar{q}\sigma_+-q\sigma_-.\nonumber
\eea
where, the Pauli matrices satisfy the $SU(2)$ algebra:

\be
[\sigma_+,\sigma_-]=\sigma_3\quad{\rm and}\quad [\sigma_3,\sigma_\pm]=\pm 2\sigma_\pm,\label{4}
\ee
and $q,~\bar{q}$ are mutually conjugate amplitudes, this leads to the coupled KdV equations. It is easy to see that
an $sl(2)$-loop algebra can be constructed on the $SU(2)$ basis, which in turn enables a complete gauge-group interpretation of this system.
\paragraph*{}Incorporating the generic representation of Eq.s \ref{3N}, the Lax pair takes the form,

\bea
&&A=\bar{q}\sigma_+-q\sigma_- \quad{\rm and}\nonumber\\
&&B=\left(\bar{q}q_x-q\bar{q}_x\right)\sigma_3-\left(\bar{q}_{xx}+2q\bar{q}^2\right)\sigma_++\left(q_{xx}+2\bar{q}q^2\right)\sigma_-.\label{23}
\eea
The corresponding curvature then can be evaluated as,

\be
F_{tx}=\left(\bar{q}_t+\bar{q}_{xxx}+6q\bar{q}\bar{q}_x\right)\sigma_+-\left(q_t+q_{xxx}+6q\bar{q}q_x\right)\sigma_-,\label{24}
\ee
yielding two coupled KdV-like equations,

\be
\bar{q}_t+\bar{q}_{xxx}+6q\bar{q}\bar{q}_x=0 \quad{\rm and}\quad q_t+q_{xxx}+6q\bar{q}q_x=0,\label{25}
\ee
under zero-curvature condition, by considering each linearly independent component of the curvature matrix. Although the
above equations posses higher order non-linearity than the usual KdV system, a straight-forward choice of variables, 
\be
\bar{q}(q)=1, \quad{\rm and}\quad q(\bar{q})=u;\qquad u\in{\mathbb R},\label{5}
\ee
immediately leads to the {\it non-coupled} (usual) KdV equation,

\be
u_t+6uu_x+u_{xxx}=0.\label{8}
\ee
The other possibility: $\bar{q}(q)=-u,~q(\bar{q})=1$ leads to a KdV equation with a negative sign to the non-linear
term, which can be transformed to the `usual' one through the transformation $u\rightarrow -u$. The Lax pair corresponding
to the choice in Eq. \ref{5} is,
\bea
&&A=\sigma_+-u\sigma_- \quad{\rm and}\nonumber\\
&&B=u_x\sigma_3-2u\sigma_++\left(u_{xx}+2u^2\right)\sigma_-,\label{6}
\eea
leading to the curvature,

\be
F_{tx}:=A_t-B_x+[A,B]\equiv-\left(u_t+6uu_x+u_{xxx}\right)\sigma_-,\label{7}
\ee
which vanishes {\it on-shell} subjected to the KdV equation. This algebraic structure stemming from the $SU(2)$ representation allows construction of the $sl(2)$ loop algebra necessary for the Abelianization procedure of quasi-integrability \cite{FZ,FZ1,FZ2}. This would not have been possible with the more common monic Lax pair,

\be
A=\partial_x+\partial_x^2+u,\quad{\rm and}\quad B=-4\partial_x^3-6u\partial_x-3u_x,
\ee
for the KdV equation. In the following we explicate the Abelianization procedure in detail.

\subsection{Quasi-integrable Deformation}\label{SS2.2}
Since the KdV equation has derivatives higher than two, a dynamic interpretation 
of the same is not possible at the level of the equation itself. In order to employ the quasi-integrability mechanism of 
Ref.s \cite{FZ,FZ1,FZ2,N1}, the notion of {\it potential} is essential, that emerges from such an interpretation of the equation.
In case of KdV system, however, a well-known Hamiltonian formulation \cite{2} exists. In fact, the KdV equation \ref{8}
can be shown to emerge from two different equivalent Hamiltonians. Subjected to the order of non-linearity appearing in
the Lax pair of Eq. \ref{6}, we opt for the following Hamiltonian,

\be
H_1[u]=\int_{-\infty}^\infty dx~\left(\frac{1}{2}u_x^2-u^3\right) \quad{\rm with}\quad \frac{\delta H_1[u]}{\delta u(x)}=-3u^2-u_{xx}.\label{9}
\ee
This enables us to re-express the temporal Lax component ($B$) of the Lax pair as,

\be
B\equiv u_x\sigma_3-2u\sigma_++\left[u_{xx}-\frac{2}{3}\left(\frac{\delta H_1[u]}{\delta u(x)}+u_{xx}\right)\right]\sigma_-.\label{11}
\ee 
The above is a general expression to accommodate any possible deformation at the Hamiltonian level. We propose that the deformation of the system is implemented in the non-linear part of Hamiltonian for the KdV system to impart
quasi-integrability, the explicit form of which will be discussed below. The corresponding curvature takes the form:

\be
F_{tx}\equiv\left[u_t+u_{xxx}-\frac{2}{3}\partial_x\left(\frac{\delta H_1[u]}{\delta u(x)}+u_{xx}\right)+2uu_x\right]\sigma_++{\cal X}\sigma_3,\label{12}
\ee
with the supposed anomaly term,

\be
{\cal X}=2u^2+\frac{2}{3}\left(\frac{\delta H_1[u]}{\delta u(x)}+u_{xx}\right),\label{13}
\ee
that vanishes for undeformed system\footnote{\noindent One can very well work with the {\it second} Hamiltonian form for the KdV
system \cite{2}: $H_2[u]=-\int_{-\infty}^\infty dx~\frac{1}{2}u^2(x)$, with the alternate fundamental bracket defined as
$\left\{u(x),u(y)\right\}=\left[\partial^3_x+2\left(u_x+u\partial_x\right)\right]\delta(x-y)$. Then, the time component 
of the Lax pair will take the form: $B\equiv-u_x\sigma_3-\left[u_{xx}-4\frac{\delta H_2[u]}{\delta x}\right]\sigma_++2u\sigma_-$.
Rest will follow through the replacement: 
$\frac{2}{3}\left(\frac{\delta H_1[u]}{\delta u(x)}+u_{xx}\right)\rightarrow 4\frac{\delta H_2[u]}{\delta x}$.}. In
the presence of this anomaly, implementation of the {\it deformed} `equation of motion' (EOM) ({\it i. e.}, the KdV equation),

\bea
&&u_t+u_{xxx}-\frac{2}{3}\partial_x\left(\frac{\delta H_1[u]}{\delta u(x)}+u_{xx}\right)+2uu_x=0,\nonumber\\
{\rm or},\quad&&u_t+6uu_x+u_{xxx}={\cal X}_x,\label{14}
\eea
leaves the curvature non-zero. 
\paragraph*{}Starting from the deformed Lax pair, one can construct an infinite number of quasi-conserved charges through the Abelianization procedure applied
in Ref.s \cite{FZ,FZ1,FZ2}, through gauge-transforming the Lax components:
\be
(A,B)\rightarrow U(A,B)U^{-1}+U_{(x,t)}U^{-1}\implies F_{tx}\rightarrow UF_{tx}U^{-1},\label{15}
\ee
In doing so, the anomaly ${\cal X}$ prevents
rotation of {\it both} of them into the same infinite dimensional Abelian subalgebra of the characteristic $sl(2)$ loop
algebra, eventually leading to an infinite set of quasi-conservation laws characterized by ${\cal X}$. 
\paragraph*{\it The $sl(2)$ loop algebra:}The $SU(2)$ algebraic structure for the KdV system \cite{1} enables the construction of an $sl(2)$ loop algebra:

\be
\left[F^m,F_\pm^m\right]=2F_\mp^{m+n}, \quad \left[F_-^m,F_+^n\right]=F^{m+n+1},\label{19}
\ee
consistent with the definitions,

\be
F^n=\lambda^n\sigma_3, \quad F_-^n=\frac{\lambda^n}{\sqrt{2}}\left(\sigma_+-\lambda\sigma_-\right) \quad{\rm and}\quad F_+^n=\frac{\lambda^n}{\sqrt{2}}\left(\sigma_++\lambda\sigma_-\right),\label{20}
\ee
with $\lambda$ being the spectral parameter. Such a structure is essentially same as that in Ref. \cite{FZ2} for quasi-integrable (QI) NLS systems. This serves as a strong connection between the quasi-deformations of the two systems, which we will address soon. 

\paragraph*{\it The Gauge Transformation:} The Lax pair of Eq. \ref{6} with $B$ deformed according to Eq. \ref{11}, however, is {\it not} suitable for the Abelianization (a version of the Drinfeld-Sokolov reduction) as the the spatial component $A$ does not contain a constant semi-simple element of the $sl(2)$ algebra that split the algebra into the correspondingKernel and Image subspaces.
In other words, this mandates the presence of the spectral parameter ($\lambda$) in the Lax pair in a particular way. A Lax pair that fulfills this algebraic requirement and also leads to the quasi-KdV equation is \cite{NN001},

\bea
&&\bar{A}=\sigma_+-(u-\lambda)\sigma_- \quad{\rm and}\nonumber\\
&&\bar{B}=u_x\sigma_3-(2u+4\lambda)\sigma_++\left\{u_{xx}-\frac{2}{3}\left(\frac{\delta H_1[u]}{\delta u}+u_{xx}\right)+2\lambda u-4\lambda^2\right\}\sigma_-,\label{N01}
\eea
with $B$ being suitably quasi-deformed which we will adopt for the remaining analysis. It is clear that the spatial Lax operator is free from the quasi-deformation, a crucial property exploited by the Abelianization procedure to obtain the general form of the quasi-conserved charges. The undeformed version of the above Lax pair can be obtained from the previous one through a gauge transformation corresponding to a unitary operator\footnote{It might not be the most general gauge transformation that leads to the desired Lax pair. Further the time-dependence of $a_\pm$ may include some non-trivial extensions. As the particular form in $\bar{A}$ is needed, the corresponding undeformed $\bar{B}$ may suitably constructed through term-by-term compensation starting with $B$.},
 
 \bea
 &&{\cal G}=\exp\left(a_+\sigma_++a_-\sigma_-\right);\nonumber\\
 &&a_+=\frac{\lambda}{\sqrt{\lambda-2u}}\int_x\frac{1}{\sqrt{\lambda-2u}},~a_-=\frac{\lambda}{2}\sqrt{\lambda-2u}\int_x\frac{1}{\sqrt{\lambda-2u}}.\label{2N13}
 \eea
 In terms of the $sl(2)$ generators the new Lax operators take the forms,
 
\bea
&&\bar{A}=\sqrt{2}F_+^0-\frac{u}{\sqrt{2}}\left(F_+^{-1}-F_-^{-1}\right) \quad{\rm and}\nonumber\\
&&\bar{B}=-4\sqrt{2}F_+^1+u_xF^0-2\sqrt{2}uF_+^0+\frac{f(u)}{\sqrt{2}}\left(F_+^{-1}-F_-^{-1}\right)\label{N02}\\
&&{\rm where}\quad f(u)=\left\{u_{xx}-\frac{2}{3}\left(\frac{\delta H_1[u]}{\delta u}\right)+u_{xx}\right\}\nonumber,
\eea
which are essentially similar to those given in Ref. \cite{KdVQI1} subjected to a particular interpretation of the deformed Hamiltonian $H_1[u]$. The above structure incorporates all possibilities of anomalous deformation of the KdV equations and therefore should correspond to multiple quasi-KdV systems.
\paragraph*{}Following the general approach of in the Ref.s \cite{FZ,FZ1,FZ2} for Abelianization by gauge-rotating the spatial Lax operator exclusively to the image of $sl(2)$, we undertake the gauge transformation defined by,

\be
g=e^G,\qquad G=\sum_{n=-1}^{-\infty}\alpha_nF_-^n+\beta_nF^n.\label{21}
\ee 
Here the coefficients $\alpha_{-n},~\beta_{-n}$ are to be chosen such that the transformed spatial component $\tilde{A}=g\bar{A}g^{-1}+g_xg^{-1}$ depends only on $F_+^n$s:
\be
\tilde{A}\equiv\sum_{n=0}^{-\infty}\gamma_nF_+^n.\label{22}
\ee 
On employing the BCH formula $e^XYe^{-X}=Y+[X,Y]+\frac{1}{2!}[X,[X,Y]]+\frac{1}{3!}[X,[X,[X,Y]]]+\cdots$ the new spatial component has the general form,

\bea
&&\tilde{A}=\bar{A}+[G,\bar{A}]+\frac{1}{2!}[G,[G,\bar{A}]]+\frac{1}{3!}[G,[G,[G,\bar{A}]]]+\cdots\nonumber\\
&&\qquad+G_x+\frac{1}{2!}[G,G_x]+\frac{1}{3!}[G,[G,G_x]]+\cdots,\label{N1}
\eea
The first few of the individual commutators are,

\bea
&&[G,\bar{A}]=\sum_{n=-1}^{-\infty}\left[\sqrt{2}\alpha_n\left(F^{n+1}-\frac{u}{2}F^n\right)+\sqrt{2}\beta_n\left(2F_-^n-u\beta_nF_-^{n-1}+u\beta_nF_+^{n-1}\right)\right],\nonumber\\
&&\frac{1}{2!}[G,[G,\bar{A}]]\nonumber\\
&&=\sqrt{2}\sum_{m,n=-1}^{-\infty}\Bigg[\left(\frac{u}{2}\alpha_m\alpha_n+2\beta_m\beta_n\right)F_+^{m+n}+\frac{u}{2}\alpha_m\beta_nF^{m+n}-\alpha_m\alpha_nF_+^{m+n+1}\nonumber\\
&&\qquad-u\beta_m\beta_nF_+^{m+n-1}+u\beta_m\beta_nF_-^{m+n-1}\Bigg],\nonumber\\
&&\frac{1}{3!}[G,[G,[G,\bar{A}]]]\nonumber\\
&&=\frac{\sqrt{2}}{3}\sum_{l,m,n=-1}^{-\infty}\Bigg[-\alpha_l\alpha_m\alpha_nF^{l+m+n+2}+\alpha_l\left(\frac{u}{2}\alpha_m\alpha_n+2\beta_m\beta_n\right)F^{l+m+n+1}\nonumber\\
&&\qquad-u\alpha_l\beta_m\beta_nF^{l+m+n}-u\alpha_l\alpha_m\beta_nF_+^{m+n+l}+2u\beta_l\beta_m\beta_nF_+^{l+m+n-1}\nonumber\\
&&\qquad-2\beta_l\alpha_m\alpha_nF_-^{l+m+n+1}+\beta_l\left(u\alpha_m\alpha_n+4\beta_m\beta_n\right)F_-^{l+m+n}-2u\beta_l\beta_m\beta_nF_-^{l+m+n-1}\Bigg],\nonumber
\eea
\bea
&&\frac{1}{4!}[G,[G,[G,[G,\bar{A}]]]]\nonumber\\
&&=\frac{1}{6\sqrt{2}}\sum_{k,l,m,n=-1}^{-\infty}\Bigg[-u\alpha_l\alpha_l\alpha_m\beta_nF^{k+l+m+n+1}+2u\alpha_k\beta_l\beta_m\beta_nF^{k+l+m+n}\nonumber\\
&&\qquad+2\alpha_k\alpha_l\alpha_m\alpha_nF_+^{k+l+m+n+2}-\left\{\alpha_k\alpha_l\left(u\alpha_m\alpha_n+4\beta_m\beta_n\right)+4\beta_k\beta_l\alpha_m\alpha_n\right\}F_+^{k+l+m+n+1}\nonumber\\
&&\qquad+2\left\{u\alpha_k\alpha_l\beta_m\beta_n+\beta_k\beta_l\left(u\alpha_m\alpha_n+4\beta_m\beta_n\right)\right\}F_+^{k+l+m+n}\nonumber\\
&&\qquad-4u\beta_k\beta_l\beta_m\beta_nF_+^{k+l+m+n-1}-2u\beta_k\alpha_l\alpha_m\beta_nF_-^{k+l+m+n}+4u\beta_k\beta_l\beta_m\beta_nF_-^{k+l+m+n-1}\Bigg],\nonumber\\
&&\cdots\nonumber\\
&&{\rm and}\nonumber\\
&&\frac{1}{2!}[G,G_x]=\sum_{m,n=-1}^{-\infty}\left(\beta_m\alpha_{n,x}-\alpha_m\beta_{n,x}\right)F_+^{m+n},\nonumber\\
&&\frac{1}{3!}[G,[G,G_x]]=\frac{1}{3}\sum_{l,m,n=-1}^{-\infty}\left(\beta_m\alpha_{n,x}-\alpha_m\beta_{n,x}\right)\left(\alpha_lF^{l+m+n+1}+2\beta_lF_-^{l+m+n}\right),\nonumber\\
&&\frac{1}{4!}[G,[G,[G,G_x]]]=\frac{1}{6}\sum_{k,l,m,n=-1}^{-\infty}\left(\beta_m\alpha_{n,x}-\alpha_m\beta_{n,x}\right)\Big(2\beta_k\beta_lF_+^{k+l+m+n}\nonumber\\
&&\qquad\qquad\qquad\qquad\qquad\qquad\qquad\qquad-\alpha_k\alpha_lF_+^{k+l+m+n+1}\Big),\nonumber\\
&&\cdots\label{N2}
\eea
The order-by-order conditions of vanishing the coefficients of $F^n,F_-^n$s lead to the expressions for the expansion coefficients of the gauge operator $g$ as,

\bea
&&{\cal O}\left(F^0\right):\quad~~ \alpha_{-1}=0,\nonumber\\
&&{\cal O}\left(F^{-1}\right):\quad \alpha_{-2}=\frac{u_x}{4\sqrt{2}},\nonumber\\
&&{\cal O}\left(F^{-2}\right):\quad \alpha_{-3}=\frac{u_{xxx}}{16\sqrt{2}}+\frac{3}{8\sqrt{2}}uu_x,\nonumber\\
&&{\cal O}\left(F^{-3}\right):\quad \alpha_{-4}=\frac{u_{xxxxx}}{64\sqrt{2}}+\frac{5}{16\sqrt{2}}u_xu_{xx}+\frac{5}{32\sqrt{2}}uu_{xxx}+\frac{11}{24\sqrt{2}}u^2u_x,\nonumber\\
&&\cdots\nonumber\\
&&{\cal O}\left(F_-^{-1}\right):\quad \beta_{-1}=-\frac{u}{4},\nonumber\\
&&{\cal O}\left(F_-^{-2}\right):\quad \beta_{-2}=-\frac{u_{xx}}{16}-\frac{u^2}{8},\nonumber\\
&&{\cal O}\left(F_-^{-3}\right):\quad \beta_{-3}=-\frac{u_{xxxx}}{64}-\frac{uu_{xx}}{8}-\frac{
3}{32}(u_x)^2-\frac{u^3}{12},\nonumber\\
&&{\cal O}\left(F_-^{-4}\right):\quad \beta_{-4}=-\frac{u_{xxxxxx}}{256}-\frac{5}{64}(u_{xx})^2-\frac{15}{128}u_xu_{xxx}-\frac{3}{64}uu_{xxxx}-\frac{53}{192}u(u_x)^2\nonumber\\
&&\qquad\qquad\qquad\qquad\quad-\frac{3}{16}u^2u_{xx}-\frac{u^4}{16},\nonumber\\
&&\cdots\label{N3}
\eea
These immediately lead to the expression of the rotated spatial Lax component $\tilde{A}$ as the expansion coefficients of the  in Eq. \ref{22} are completely determined,

\bea
&&\gamma_0=\sqrt{2},\nonumber\\
&&\gamma_{-1}=-\frac{u}{\sqrt{2}},\nonumber\\
&&\gamma_{-2}=-\frac{u^2}{4\sqrt{2}},\nonumber\\
&&\gamma_{-3}=\frac{1}{16\sqrt{2}}\left((u_x)^2-uu_{xx}-2u^3\right),\nonumber\\
&&\gamma_{-4}=-\frac{uu_{xxxx}}{64\sqrt{2}}-\frac{3}{32\sqrt{2}}u(u_x)^2-\frac{u^2u_{xx}}{8\sqrt{2}}-\frac{15}{192\sqrt{2}}u^4,\nonumber\\
&&\cdots\label{N4}
\eea
in terms of the deformed solution $u$.

\paragraph*{}Subsequently, the temporal Lax component $\bar{B}$ transforms to,

\bea
&&\tilde{B}=g\bar{B}g^{-1}+g_tg^{-1}\nonumber\\
&&\quad=\bar{B}+[G,\bar{B}]+\frac{1}{2!}[G,[G,\bar{B}]]+\frac{1}{3!}[G,[G,[G,\bar{B}]]]+\cdots\nonumber\\
&&\quad\qquad+G_t+\frac{1}{2!}[G,G_t]+\frac{1}{3!}[G,[G,G_t]]+\cdots,\label{N03}
\eea
with a few of the lowest order commutators being,

\bea
&&[G,\bar{B}]\nonumber\\
&&=\sum_{n=-1}^{-\infty}\Big[-4\sqrt{2}\alpha_nF^{n+2}-2\sqrt{2}u\alpha_nF^{n+1}+\frac{f(u)}{\sqrt{2}}\alpha_nF^n-2u_x\alpha_nF_+^n-\sqrt{2}f(u)\beta_nF_+^{n-1}\nonumber\\
&&\qquad-8\sqrt{2}\beta_nF_-^{n+1}-4\sqrt{2}u\beta_nF_-^n+\sqrt{2}f(u)\beta_nF_-^{n-1}\Big]\nonumber\\
&&\frac{1}{2!}[G,[G,\bar{B}]]\nonumber\\
&&=\sum_{m,n=-1}^{-\infty}\Big[-u_x\alpha_m\alpha_nF^{m+n+1}-\frac{f(u)}{\sqrt{2}}\alpha_m\left(\alpha_n+\frac{\beta_n}{2}\right)F^{m+n}+4\sqrt{2}\alpha_m\alpha_nF_+^{m+n+2}\nonumber\\
&&\qquad+4\sqrt{2}\left(u\alpha_m\alpha_n-4\beta_m\beta_n\right)F_+^{m+n+1}-4\sqrt{2}u\beta_m\beta_nF_+^{m+n}+\sqrt{2}f(u)\beta_m\beta_nF_+^{m+n-1}\nonumber\\
&&\qquad-2u_x\beta_m\alpha_nF_-^{m+n}-\sqrt{2}f(u)\beta_m\beta_nF_-^{m+n-1}\Big],\nonumber
\eea
\bea
&&\frac{1}{3!}[G,[G,[G,\bar{B}]]]\nonumber\\
&&=\sum_{l,m,n=-1}^{-\infty}\Big[\frac{4\sqrt{2}}{3}\alpha_l\alpha_m\alpha_nF^{l+m+n+3}+\frac{2\sqrt{2}}{3}\alpha_l\left(u\alpha_m\alpha_n-4\beta_m\beta_n\right)F^{l+m+n+2}\nonumber\\
&&\qquad-\frac{4\sqrt{2}}{3}u\alpha_l\beta_m\beta_nF^{l+m+n+1}+\frac{\sqrt{2}}{3}f(u)\alpha_l\beta_m\beta_nF^{l+m+n}+\frac{2}{3}u_x\alpha_l\alpha_m\alpha_nF_+^{l+m+n+1}\nonumber\\
&&\qquad+\frac{2}{3}\left\{\frac{f(u)}{\sqrt{2}}\alpha_l\alpha_m\left(2\alpha_n-\beta_n\right)-2u_x\beta_l\beta_m\alpha_n\right\}F_+^{l+m+n}-\frac{2\sqrt{2}}{3}f(u)\beta_l\beta_m\beta_nF_+^{l+m+n-1}\nonumber\\
&&\qquad+\frac{
8\sqrt{2}}{3}\beta_l\alpha_m\alpha_nF_-^{l+m+n+2}+\frac{4\sqrt{2}}{3}\beta_l\left(u\alpha_m\alpha_n-4\beta_m\beta_n\right)F_-^{l+m+n+1}\nonumber\\
&&\qquad-\frac{8\sqrt{2}}{3}u\beta_l\beta_m\beta_nF_-^{l+m+n}+\frac{2\sqrt{2}}{3}f(u)\beta_l\beta_m\beta_nF_-^{l+m+n-1}\Big],\nonumber\\
&&\frac{1}{4!}[G,[G,[G,[G,\bar{B}]]]]\nonumber\\
&&=\sum_{k,l,m,n=-1}^{-\infty}\Big[\frac{1}{6}u_x\alpha_k\alpha_l\alpha_m\alpha_nF^{k+l+m+n+2}+\frac{1}{6}\alpha_k\left\{\frac{f(u)}{\sqrt{2}}\alpha_l\alpha_m\left(2\alpha_n-\beta_n\right)-2u_x\beta_l\beta_m\alpha_n\right\}\nonumber\\
&&\qquad\times F^{k+l+m+n+1}-\frac{f(u)}{3\sqrt{2}}\alpha_k\beta_l\beta_m\beta_nF^{k+l+m+n}-\frac{2\sqrt{2}}{3}\alpha_k\alpha_l\alpha_m\alpha_nF_+^{k+l+m+n+3}\nonumber\\
&&\qquad+\frac{\sqrt{2}}{3}\left\{4\beta_k\beta_l\alpha_m\alpha_n-\alpha_k\alpha_l\left(u\alpha_m\alpha_n-4\beta_m\beta_n\right)\right\}F_+^{k+l+m+n+2}\nonumber\\
&&\qquad+\frac{2\sqrt{2}}{3}\left\{u\alpha_k\alpha_l\beta_m\beta_n+\beta_k\beta_l\left(u\alpha_m\alpha_n-4\beta_m\beta_n\right)\right\}F_+^{k+l+m+n+1}\nonumber\\
&&\qquad-\frac{1}{3}\left\{\frac{f(u)}{\sqrt{2}}\alpha_k\alpha_l\beta_m\beta_n+4\sqrt{2}u\beta_k\beta_l\beta_m\beta_n\right\}F_+^{k+l+m+n}+\frac{\sqrt{2}f(u)}{3}\beta_k\beta_l\beta_m\beta_nF_+^{k+l+m+n-1}\nonumber\\
&&\qquad+\frac{1}{3}u_x\beta_k\alpha_l\alpha_m\alpha_nF_-^{k+l+m+n+1}+\frac{1}{3}\beta_k\left\{\frac{f(u)}{\sqrt{2}}\alpha_l\alpha_m\left(2\alpha_n-\beta_n\right)-2u_x\beta_l\beta_m\alpha_n\right\}\nonumber\\
&&\qquad\times F_-^{k+l+m+n}-\frac{\sqrt{2}}{3}f(u)\beta_k\beta_l\beta_m\beta_nF_-^{k+l+m+n-1}\Big],\nonumber\\
&&\cdots\nonumber
\eea
\bea
&&\frac{1}{2!}[G,G_t]=\sum_{m,n=-1}^{-\infty}\left(\beta_m\alpha_{n,t}-\alpha_m\beta_{n,t}\right)F_+^{m+n},\nonumber\\
&&\frac{1}{3!}[G,[G,G_t]]=\frac{1}{3}\sum_{l,m,n=-1}^{-\infty}\left(\beta_m\alpha_{n,t}-\alpha_m\beta_{n,t}\right)\left(\alpha_lF^{l+m+n+1}+2\beta_lF_-^{l+m+n}\right)\nonumber\\
&&\frac{1}{4!}[G,[G,[G,G_t]]]=\frac{1}{6}\sum_{k,l,m,n=-1}^{-\infty}\left(\beta_m\alpha_{n,t}-\alpha_m\beta_{n,t}\right)\left(2\beta_k\beta_lF_+^{k+l+m+n}-\alpha_k\alpha_lF_+^{k+l+m+n+1}\right)\nonumber\\
&&\cdots\label{N04}
\eea
The rotated temporal Lax component will span both Kernal and Image of $sl(2)$ to mandate a general form:

\be
\tilde{B}=\sum_{n=0}^{-\infty}\left[a_nF_-^n+b_nF^n+c_nF_+^n\right],\label{N6}
\ee
wherein, a few of the nontrivial expansion coefficients are,

\bea
&&a_0=2\sqrt{2}u,\nonumber\\
&&a_{-1}=-\frac{f(u)}{\sqrt{2}}+\frac{u_{xx}}{\sqrt{2}}+2\sqrt{2}u^2,\nonumber\\
&&a_{-2}=-\frac{u}{2\sqrt{2}}f(u)+\frac{u_{xt}}{4\sqrt{2}}+\frac{u_{xxx}}{4\sqrt{2}}+\frac{5}{2\sqrt{2}}uu_{xx}+\frac{3}{2\sqrt{2}}\left(u_x\right)^2+\frac{13}{6\sqrt{2}}u^3,\nonumber\\
&&a_{-3}=\frac{u_{xxxt}}{16\sqrt{2}}+\frac{3}{8\sqrt{2}}\left(u_tu_x+uu_{xt}\right)+\frac{u_{xxxxxx}}{16\sqrt{2}}+\frac{5}{4\sqrt{2}}\left(u_{xx}\right)^2+\frac{15}{8\sqrt{2}}u_xu_{xxx}+\frac{7}{8\sqrt{2}}uu_{xxxx}\nonumber\\
&&\qquad+\frac{19}{4\sqrt{2}}u\left(u_x\right)^2+\frac{33}{8\sqrt{2}}u^2u_{xx}+\frac{13}{6\sqrt{2}}u^4-\frac{f(u)}{8\sqrt{2}}\left(u_{xx}+3u^2\right),\nonumber\\
&&\cdots\label{N7}\\
&&b_1=0=b_2,\nonumber\\
&&b_{-1}=-\frac{u_t}{4}-\frac{u_{xxx}}{4}-2uu_x,\nonumber\\
&&b_{-2}=-\frac{u_{xxt}}{16}-\frac{uu_t}{4}+\frac{f(u)}{8}u_x-\frac{u_{xxxxx}}{16}-\frac{5}{4}u_xu_{xx}-\frac{3}{4}uu_{xxx}-\frac{21}{8}u^2u_x,\nonumber\\
&&\cdots\label{N8}\\
&&c_1=-4\sqrt{2},\nonumber\\
&&c_0=-2\sqrt{2}u,\nonumber\\
&&c_{-1}=\frac{1}{\sqrt{2}}\left(f(u)-u^2\right),\nonumber\\
&&c_{-2}=\frac{f(u)}{2\sqrt{2}}u-\frac{\left(u_x\right)^2}{4\sqrt{2}}-\frac{uu_{xx}}{2\sqrt{2}}-\frac{3}{2\sqrt{2}}u^3,\nonumber\\
&&\cdots\label{N9}
\eea
\paragraph*{}Preceding the gauge transformation, following Eq. \ref{N02}, the curvature has the form:

\bea
&&\bar{F}_{tx}=\bar{A}_t-\bar{B}_x+[\bar{A},\bar{B}]\nonumber\\
&&\qquad=-\left[u_t+u_{xxx}+2uu_x-\frac{2}{3}\left(\frac{\delta H_1[u]}{\delta u}+u_{xx}\right)_x\right]\sigma_--{\cal X}\sigma_3\equiv-{\cal X} F^0,\label{N05}
\eea
wherein the coefficient of $\sigma_-$ vanish following the deformed KdV equation. Then,
following the gauge transformation yields the rotated curvature as:
\bea
&&\bar{F}_{tx}\to\tilde{F}_{tx}=g\bar{F}_{tx}g^{-1}\nonumber\\
&&\qquad\qquad\quad=\bar{F}_{tx}+[G,\bar{F}_{tx}]+\frac{1}{2!}[G,[G,\bar{F}_{tx}]]+\frac{1}{3!}[G,[G,[G,\bar{F}_{tx}]]]+\cdots,\label{N06}
\eea
wherein a few of the commutators have the forms:

\bea
&&[G,\bar{F}_{tx}]=2{\cal X}\sum_{n=-1}^{-\infty}\alpha_nF_+^n,\nonumber\\
&&\frac{1}{2!}[G,[G,\bar{F}_{tx}]]={\cal X}\sum_{m,n=-1}^{-\infty}\left(\alpha_m\alpha_nF^{m+n+1}+2\beta_m\alpha_nF_-^{m+n}\right),\nonumber\\
&&\frac{1}{3!}[G,[G,[G,\bar{F}_{tx}]]]=\frac{2}{3}{\cal X}\sum_{l,m,n=-1}^{-\infty}\left(2\beta_l\beta_m\alpha_nF_+^{l+m+n}-\alpha_l\alpha_m\alpha_nF_+^{l+m+n+1}\right),\nonumber\\
&&\frac{1}{4!}[G,[G,[G,[G,\bar{F}_{tx}]]]]\nonumber\\
&&\qquad=\frac{{\cal X}}{6}\sum_{k,l,m,n=-1}^{-\infty}\Big[2\alpha_k\beta_l\beta_m\alpha_nF^{k+l+m+n+1}-\alpha_k\alpha_l\alpha_m\alpha_nF^{k+l+m+n+2}+4\beta_k\beta_l\beta_m\alpha_n\nonumber\\
&&\qquad\qquad\qquad\qquad\qquad\times F_-^{k+l+m+n}-2\beta_k\alpha_l\alpha_m\alpha_nF^{k+l+m+n+1}\Big]\nonumber\\
&&\frac{1}{5!}[G,[G,[G,[G[G,,\bar{F}_{tx}]]]]]\nonumber\\
&&\qquad=\frac{{\cal X}}{15}\sum_{j,k,l,m,n=-1}^{-\infty}\Big[\alpha_j\alpha_k\alpha_l\alpha_m\alpha_nF_+^{j+k+l+m+n+2}-2\left(\alpha_j\alpha_k\beta_l\beta_m\beta_n+\beta_j\beta_k\alpha_l\alpha_m\alpha_n\right)\nonumber\\
&&\qquad\qquad\qquad\qquad\qquad\quad\times F_+^{j+k+l+m+n+1}+4\beta_j\beta_k\beta_l\beta_m\alpha_nF_+^{j+k+l+m+n}\Big],\nonumber\\
&&\cdots\label{N07}
\eea
On the other hand, as $\bar{F}_{tx}\propto F^0$, the general form of the rotated curvature will have the form,

\be
\tilde{F}_{tx}={\cal X}\sum_n\left(f_nF^n+f_n^+F_+^n+f_n^-F_-^n\right).\label{N08}
\ee
On comparing the two expressions of the rotated curvature, one obtains a few of its nontrivial expansion coefficients in terms of the system variables as,

\bea
&&f_0=-1,\nonumber\\
&&f_{-1}=0=f_{-2},\nonumber\\
&&f_{-3}=\frac{\left(u_x\right)^2}{16},\nonumber\\
&&f_{-4}=\frac{u_{xxx}}{64}+\frac{3}{32}uu_x,\nonumber\\
&&\cdots\label{N09}
\eea
\bea
&&f^+_{-1}=0,\nonumber\\
&&f^+_{-2}=\frac{u_x}{2\sqrt{2}},\nonumber\\
&&f^+_{-3}=\frac{u_{xxx}}{8\sqrt{2}}+\frac{3}{4\sqrt{2}}uu_x,\nonumber\\
&&f^+_{-4}=\frac{u_{xxxxx}}{32\sqrt{2}}+\frac{5}{8\sqrt{2}}u_xu_{xx}+\frac{5}{16\sqrt{2}}uu_{xxx}+\frac{15}{16\sqrt{2}}u^2u_x,\nonumber\\
&&\cdots\label{N010}\\
&&f^-_{-2}=0,\nonumber\\
&&f^-_{-3}=-\frac{uu_x}{8\sqrt{2}},\nonumber\\
&&f^-_{-4}=-\frac{uu_{xxx}}{32\sqrt{2}}-\frac{u_xu_{xx}}{32\sqrt{2}}-\frac{u^2u_{xx}}{4\sqrt{2}},\nonumber\\
&&f^-_{-5}=-\frac{uu_{xxxxx}}{128\sqrt{2}}-\frac{u_xu_{xxxx}}{128\sqrt{2}}-\frac{u_{xx}u_{xxx}}{128\sqrt{2}}-\frac{15}{64\sqrt{2}}uu_xu_{xx}-\frac{3}{64\sqrt{2}}\left(u_x\right)^3-\frac{141}{384\sqrt{2}}u^3u_x\nonumber\\
&&\qquad\quad-\frac{3}{32\sqrt{2}}u^2u_{xxx},\nonumber\\
&&\cdots\label{N011}
\eea
The rotated curvature, on the other hand, can directly be obtained from the corresponding Lax pair $\left(\tilde{A},\tilde{B}\right)$ as,

\bea
&&\tilde{F}_{tx}=\tilde{A}_t-\tilde{B}_x+[\tilde{A},\tilde{B}]\nonumber\\
&&\qquad\equiv\sum_n\Big[\left(\gamma_{n,t}-c_{n,x}\right)F_+^n-\left(b_{n,x}-2\gamma_n\sum_ma_mF^{m+1}\right)F^n\nonumber\\
&&\qquad\quad-\left(a_{n,x}+2\gamma_n\sum_mb_mF_-^m\right)F_-^n\Big].\label{N012}
\eea 
It can be seen that the projection of the rotated curvature onto the Image sub-sector defined by the generators $F_+^n$, to which $\tilde{A}$ was rotated exclusively, is linear in the coefficient. This will be crucial for defining charges for this system.

\subsection{Quasi-conservation}\label{2.3}In order to demonstrate the deviation from integrability, based on the QI deformation,
it is pertinent to evaluate quantities which would have represent conservation or have themselves be conserved for the
undeformed system. The deliberate gauge transformation in Eq. \ref{21} rotates one (spatial) Lax component to an Abelian sub-algebra spanned by $\{F_+^n\}$ which is a particular Drinfeld-Sokolov reduction. This essentially isolates the corresponding contribution to the curvature $F_{tx}$ from any nonlinearity arising from the general non-commutation of the Lax components. For an undeformed system, owing to its integrability manifesting as the vanishing of the curvature at each spectral order, the contribution to the curvature from the liearized sub-algebra yields a very simple `continuity' relation: $\gamma_{n,t}-c_{n,x}=0$ following Eq. \ref{N010}. This enables one to construct conserved charges:

\be
Q^n=\int_x\gamma_n\rightarrow\frac{dQ^n}{dt}=\int_x\left(\gamma_{n,t}-c_{n,x}\right)\equiv 0,\label{N013}
\ee
subjected to feasible boundary behavior of the coefficients $c_n$ as they exclusively depend on the solution $u$ which is local for all the purposes of interest. As the curvature does not vanish for the deformed system, on comparing its two expressions in Eq.s \ref{N08} and \ref{N012}, these charges turn out to be non-conserved in general:

\be
\frac{dQ^n}{dt}=\int_x\left(\gamma_{n,t}-c_{n,x}\right)=\int_x{\cal X} f_n^+\equiv\Gamma^n,\label{N014}
\ee
exclusively because the anomaly ${\cal X}$ is non-zero. However, a subset of them can still be conserved depending on particular values of the expansion coefficients $f_n^+$. For example, trivially, the charge $Q^{-1}$ is identically conserved following Eq.s \ref{N010}\footnote{This particular conservation is essentially a statement of the deformed KdV equation being satisfied, as one can check by substituting for $\gamma_{-1}$ and $c_{-1}$. This further serves as a testament to the locality of $u$, which was assumed for its vanishing on the spatial boundary, as $Q_{-1}$ needs to be a constant. Additionally it may be related to the conserved energy of the system \cite{KdVQI1} as observed earlier in Ref.s \cite{FZ,FZ1,FZ2}.}. In  general, for a given deformed solution $u$, the coefficients $\gamma_n$ can be well-localized for a particular set $\{n\}$ resulting in a constant value for the corresponding $Q^n$s. Finally the anomaly ${\cal X}$ itself can have certain overall symmetry which will yield a vanishing derivative for $Q^n$s for a particular subset $\{f_n^+\}$, a topic that will be illustrated upon in the next section. All these possibilities may render the system quasi-integrable having a subset of conserved charges, since being functions of a localized deformed solution $u$ of the system they are expected to vanish asymptotically.
\paragraph*{}As discussed in Ref. \cite{KdVQI1}, though for a particular kind of quasi-deformations, the anomalous charges regain conservation for the deformed solution $u$ being well-localized; either solitonic or even multi-solitonic but well-separated. However, such solutions were subjected to a full numerical treatment which is beyond the scope of the present work. Instead we focus on various particular forms of the deformed Hamiltonian $H_1[u]$ leading to classes of possible quasi-KdV systems, which are explicated in the next section.

\section{General Quasi-integrability of The KdV System and Possible Deformations}\label{Sec3}
In order to explore the details for quasi-integrability of the KdV system, we now utilize the $\mathbb{Z}_2$ symmetry of $sl(2)$ loop algebra, which has strong similarity with that corresponding to the quasi-NLS system \cite{FZ2} and agrees with the particular quasi-KdV systems in Re. \cite{KdVQI1}. It has been found that the anomaly function and the relevant expansion coefficients must posses definite space-time parities for quasi-conservation of the charges. Although the exact reason behind this is not known clearly, it might closely be related to the Abelianization approach itself. The $\mathbb{Z}_2$ transformation is a combination of the order 2 automorphism of $sl(2)$ loop algebra:

\be
\Sigma(F^n)=F^n,\quad\Sigma(F_-^n)=-F_-^n\quad{\rm and}\quad\Sigma(F_+^n)=-F_+^n,\label{N33}
\ee
and parity:

\be
{\cal P}: \quad (\tilde{x},\tilde{t})\to(-\tilde{x},-\tilde{t});\quad{\rm with}\quad \tilde{x}=x-x_0 \quad{\rm and}\quad \tilde{t}=t-t_0
\ee
about a particular point $(x_0,t_0)$ in space-time, which can very well be chosen to be the origin. These transformations mutually commute, as they work in two different spaces ({\it i. e.}, group and coordinate subspaces). Thus,

\be
\Omega\left(\bar{A}\right)=-\bar{A},\quad\Omega=\Sigma{\cal P},\label{N34}
\ee
for $u$ being parity-even, which makes sense as we are interested in localized (solitonic) structures\footnote{One can very well identify ($x_0$,~$t_0$) as the centre of such a solitonic structure.}. The KdV equation is parity-invariant to begin with, and its quasi-modification (Eq. \ref{14}) is also the same, subjected to the explicit deformation(s) to be introduced in the next section\footnote{Practically it amounts to having ${\cal X}_x$ odd in derivatives, which it is.}. More intuitively, as the quasi-deformed systems are known to support single-soliton structures similar to those of the undeformed systems \cite{FZ,FZ1,FZ2,KdVQI1} and since the well-known bright and dark KdV solitons are parity-even, it is sensible to expect that $u$ could be such \cite{KdVQI1}. 
\paragraph*{}The use of $\mathbb{Z}_2$ transformation felicitate the asymptotic vanishing of the integral of ${\cal X} f_n^+$s in a general way, thereby ensuring conservation of the corresponding charges $Q^n$s \cite{FZ2}. For this purpose the generator $F_+^n$\footnote{For the given Lax pair in Eq. \ref{N01}.} serves as the semi-simple element that splits the $sl(2)$ loop algebra into two sub-sectors. We have already achieved it in subsection \ref{SS2.2} through the gauge transformation, a general version of which can be expressed as,

\be
\mathfrak g=\exp\sum_{n=-1}^{-\infty}\mathfrak G^n,\label{N35}
\ee
where $\mathfrak G^n$ is any linear combination of the generators $F^n,~F_-^n$ that rotates to the sub-sector defined by $F_+^n$. Then the gauge-rotated spatial connection, takes the form,

\be
\bar{A}\to\tilde{\cal A}={\mathfrak g}\bar{A}{\mathfrak g}^{-1}+{\mathfrak g}_x{\mathfrak g}^{-1},\label{N36}
\ee
which continues to be a eigenstate of $\Omega$. To see this, considering the BCH expansions as in Eq. \ref{N1}, we can identify contributions to $\tilde{\cal A}=\sum_n\tilde{\cal A}_n$ for different $n$ (powers of spectral parameter $\lambda$) as,

\bea
&&\tilde{\cal A}_0=\bar{A}_0,\nonumber\\
&&\tilde{\cal A}_{-1}=\left[\mathfrak G^{-1},\bar{A}_0\right]+\bar{A}_{-1}+\mathfrak G^{-1}_x,\nonumber\\
&&\tilde{\cal A}_{-2}=\left[\mathfrak G^{-2},\bar{A}_0\right]+\left[\mathfrak G^{-1},\bar{A}_{-1}\right]+\frac{1}{2!}\left[\mathfrak G^{-1},\left[\mathfrak G^{-1},\bar{A}_0\right]\right]+\mathfrak G^{-2}_x+\frac{1}{2!}\left[\mathfrak G^{-1},\mathfrak G^{-1}_x\right],\nonumber\\
&&\tilde{\cal A}_{-3}=\left[\mathfrak G^{-3},\bar{A}_0\right]+\left[\mathfrak G^{-2},\bar{A}_{-1}\right]+\frac{1}{2!}\Big(\left[\mathfrak G^{-2},\left[\mathfrak G^{-1},\bar{A}_0\right]\right]+\left[\mathfrak G^{-1},\left[\mathfrak G^{-2},\bar{A}_0\right]\right]\Big)\nonumber\\
&&\qquad\quad+\frac{1}{2!}\left[\mathfrak G^{-1},\left[\mathfrak G^{-1},\bar{A}_{-1}\right]\right]+\frac{1}{3!}\left[\mathfrak G^{-1},\left[\mathfrak G^{-1},\left[\mathfrak G^{-1},\bar{A}_0\right]\right]\right]+\mathfrak G^{-3}_x+\frac{1}{2!}\Big(\left[\mathfrak G^{-2},\mathfrak G^{-1}_x\right]\nonumber\\
&&\qquad\quad+\left[\mathfrak G^{-1},\mathfrak G^{-2}_x\right]\Big)+\frac{1}{3!}\left[\mathfrak G^{-1},\left[\mathfrak G^{-1},\mathfrak G^{-1}_x\right]\right],\nonumber\\
&&\qquad\quad\vdots\label{N37}\\
&&{\rm where}\quad \bar{A}=\bar{A}_0+\bar{A}_{-1},\qquad \bar{A}_0=\sqrt{2}F_+^0,\quad \bar{A}_{-1}=-\frac{u}{\sqrt{2}}\left(F_+^{-1}-F_-^{-1}\right).\nonumber
\eea 
Wherein terms with numerical prefixes are separated according to their individual grades (powers of $\lambda$). We can immediately conclude that $\Omega\left(\tilde{\cal A}_0\right)=-\tilde{\cal A}_0$. From the second equation,

\be
\Omega\left(\tilde{\cal A}_{-1}\right)\equiv-\left[\Omega\left(\mathfrak G^{-1}\right),\bar{A}_0\right]-\bar{A}_{-1}+\Omega\left(\mathfrak G^{-1}_x\right),\label{2N01}
\ee
which when added back to the second equation yields,

\be
(1+\Omega)\left(\tilde{\cal A}_{-1}\right)=\left[(1-\Omega)\left(\mathfrak G^{-1}\right),\bar{A}_0-\partial_x\right].\label{2N02}
\ee
In the above, the LHS is exclusively in the $sl(2)$ sub-sector defined by $F_+^n$ whereas the RHS is excluded of it. Thus both the sides identically vanishes, with the RHS non-trivially leading to $\Omega\left(\mathfrak G^{-1}\right)=\mathfrak G^{-1}$. One can similarly proceed from the third of Eq.s \ref{N37} onward to obtain $\Omega\left(\mathfrak G^n\right)=\mathfrak G^n~\forall~n\in\mathbb{Z}_-$, eventually leading to $\Omega\left(\mathfrak g\right)=\mathfrak g$. This finally implies $\Omega\left(\tilde{\cal A}\right)=-\tilde{\cal A}$ from Eq. \ref{N36}.

\paragraph*{}Therefore the $\mathbb{Z}_2$ automorphism is preserved for the spatial connection under the Abelianising gauge transformation. One can check this explicitly from the expansion coefficients obtained for the particular case of $\tilde{A}$ in the last section, which is manifested through their parity properties as,

\be
{\cal P}\left(\alpha_n\right)=-\alpha_n\quad{\rm and}\quad{\cal P}\left(\beta_n\right)=\beta_n.\label{2N03}
\ee
This eventually implies definite parity properties of the coefficients $f_n^+$ of the rotated curvature in the linearized sub-sector defined by $F_+^n$s. To see this we utilize the Killing form of the $sl(2)$ loop algebra \cite{FZ2}:

\bea
&&\mathfrak K\left(F^nF^m\right)=2\delta_{m+n,0},\quad \mathfrak K\left(F_\pm^nF_\pm^m\right)=2\delta_{m+n+1,0},\quad \mathfrak K\left(F^nF_\pm^m\right)=0=\mathfrak K\left(F_+^nF_-^m\right);\nonumber\\
&&{\rm wherein}\quad \mathfrak K(\bullet)=-\frac{i}{2\pi}\oint\frac{d\lambda}{\lambda}Tr(\bullet).\label{2N04}
\eea
Then from Eq.s \ref{N05} and \ref{N08} one can express the expansion coefficients of $\tilde{F}_{tx}$ in the linearized sub-sector as,

\be
f_n^+\equiv\frac{1}{2}\mathfrak K\left(-gF^0g^{-1}F_+^{-n-1}\right)\quad\forall~n\in\mathbb{Z}_+.\label{2N05}
\ee
Since the Killing form is invariant under $\Omega$, 

\be
{\cal P}\left(f_n^+\right)\equiv\Omega\left(f_n^+\right)=\frac{1}{2}\mathfrak K\left(-\Omega(g)\Omega\left(F^0\right)\Omega\left(g^{-1}\right)\Omega\left(F_+^{-n-1}\right)\right)\equiv-f_n^+.\label{2N06}
\ee
One can explicitly check this to be true from the particular expressions obtained for $f_n^+$s in the previous section. Therefore, from Eq. \ref{N014},

\be
Q^n(t=\tilde{t})-Q^n(t=-\tilde{t})=\int_{-\tilde{t}}^{\tilde{t}}\int_{-\tilde{x}}^{\tilde{x}}{\cal X} f_n^+\equiv 0,\label{2N07}
\ee
for the anomaly function ${\cal X}$ being parity-even. In the above, $\pm(\tilde{x},\tilde{t})$ refers to spatiotemporal infinity where all the charges are supposed to vanish for sensible quasi-integrability, thereby ensuring the same for the deformation under consideration. This general parity properties of the system agrees in detail with those obtained in Ref. \cite{KdVQI1} for the RLW and mRLW systems. Therein analytical derivation and numerical evolution of one-, two- and three-soliton solutions of these quasi-KdV systems display the parity-evenness distinctly. Apart from when they were interacting the corresponding anomaly and expansion coefficients display the exact parity properties obtained here in order to conserve the charges. Following the close relation between the parity property and quasi-integrability of various systems \cite{FZ,FZ1,FZ2,N1,Own3} including KdV-like systems \cite{KdVQI1} the preceding treatment strengthens the validity of our general quasi-deformation approach to the KdV-system. 
\paragraph*{}For the quasi-NLS system \cite{FZ2} the anomaly needed to be parity-odd. For the quasi-KdV it is crucial to obtain parity-even ${\cal X}$ instead, which essentially means the Hamiltonian $H_1[u]$ needs to be modified judiciously. The undeformed Hamiltonian in Eq. \ref{9} is parity-even. Thus, from the expression of the anomaly (Eq. \ref{13}), we need a parity-even extension to $H_1[u]$ to obtain a parity-even ${\cal X}$. For example, with a deformation of the form,

\be
H_1[u]\to H_1[u]=\int_{-\infty}^\infty\left[\frac{1}{2}u_x^2-u^3+\epsilon F(u)\right],\quad\epsilon\in\mathbb{N},\label{N38}
\ee
where $F(u)=\frac{3}{4}uu_{xx}$ we obtain ${\cal X}=\epsilon u_{xx}$ which is parity-even as required. In particular, this will ensure conservation of $Q^{-2}$ along with the trivially conserved $Q^{-1}$ {\it locally} and asymptotic conservation of all $Q^n$s in general given $u$ is parity-even. To the latter end, the corresponding deformed KdV equation looks like,

\be
u_t+6uu_x+(1-\epsilon)u_{xxx}=0,\label{N39}
\ee
which essentially is a scaling of the undeformed system and thereby is integrable. Though this particular one is a somewhat trivial deformation, it is to be noted that the proposed scheme for quasi-deformation can lead to integrable deformations as a subclass. Moreover, some quasi-deformed models may asymptotically go to a scaled version of the undeformed model instead of the exact one\footnote{Or even to a non-holonomic version as observed for the NLS system \cite{AGM}.}. This equation supports single-soliton solutions of the form, 

\be
u=\frac{c}{2}{\rm sech}^2\left[\sqrt{\frac{c}{4(1-\epsilon)}}\left(x-ct-x_0+ct_0\right)\right],\quad c>0,\label{Sol1}
\ee
moving with speed $c$. This is expected as the choice for $F(u)$ is nothing but a total derivative away from the first term in $H_1[u]$. Non-trivially and more importantly, however, this provides an opportunity to construct a hierarchy of 
{\it higher-order/degree} extensions of KdV, with different choices of $F(u)$. For demonstration, we consider the following two:

\be
F(u)=\frac{3}{2m(m-1)}u^m \quad{\rm and}\quad F(u)=\frac{3}{4}uu^{(2n)}, \quad{\rm with}\quad m\in\mathbb{Z}\cdots;\quad n=1,2,\cdots,\label{N40}
\ee
where $m$ is ordinary power and $n$ is the order of space derivatives, leading to the higher-derivative equations,

\bea
&&u_t+6uu_x+u_{xxx}=\epsilon u^{m-2}u_x\quad{\rm and}\nonumber\\
&&u_t+6uu_x+u_{xxx}=\epsilon u^{(2n+1)},\label{N41}
\eea
respectively. In particular, for $n=1$ we obtain the system in Eq. \ref{N39}.
It may be worthwhile to study such extensions (deformations) of the KdV system which could admit more complicated solitonic structures than that in Eq. \ref{Sol1}. It should be pointed out that not all of them could be quasi-integrable; beyond the obvious parity-count it should be crucial that such system posses localized solutions with proper asymptotic properties as explained above.
\paragraph*{} One prospect is to deform $H_1[u]$ in such a way ${\cal X}$ forms a total derivative when multiplied with $f_+^n$s. This will automatically ensure quasi-conservation for $u$ being localized; but this may not be possible for multiple orders $n$. Such a system may not support localized solutions at all. More importantly the deformation part may not be linearly isolated in the Hamiltonian, like the `potential deformations' in Ref.s \cite{FZ,FZ1,FZ2}. Then to identify ${\cal X}$ a order-by-order approach needs to be adopted which, however, does not mean that $\epsilon$ needs to be small \cite{FZ2,AGM,Own3}.
In the next section we consider this approach utilizing the weak-coupling mapping to NLS system.

\section{Perturbative QI Deformation: The NLS Analogy}\label{Sec4}
\begin{figure}
\begin{subfigure}[t]{0.4\textwidth}
    \includegraphics[width=\textwidth]{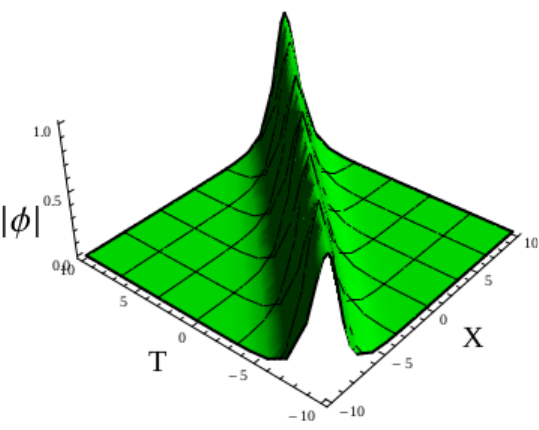}
    \caption{Undeformed single NLS soliton.}
\label{F1a}
\end{subfigure}\hspace{\fill} 
\begin{subfigure}[t]{0.5\textwidth}
    \includegraphics[width=\linewidth]{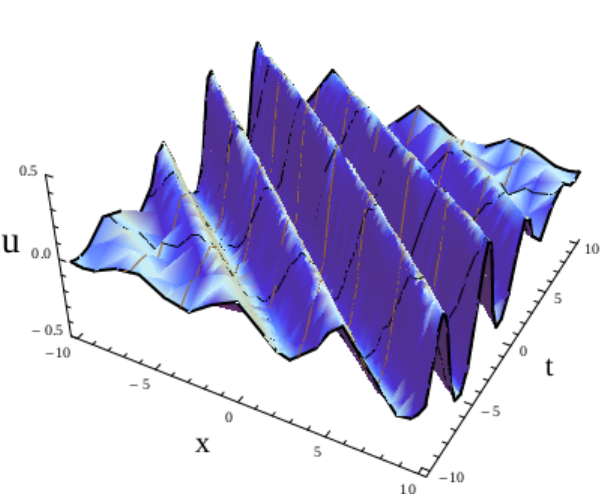}
    \caption{Undeformed mapped KdV soliton train.}
\label{F1b}
\end{subfigure}
\begin{subfigure}[t]{0.4\textwidth}
    \includegraphics[width=\linewidth]{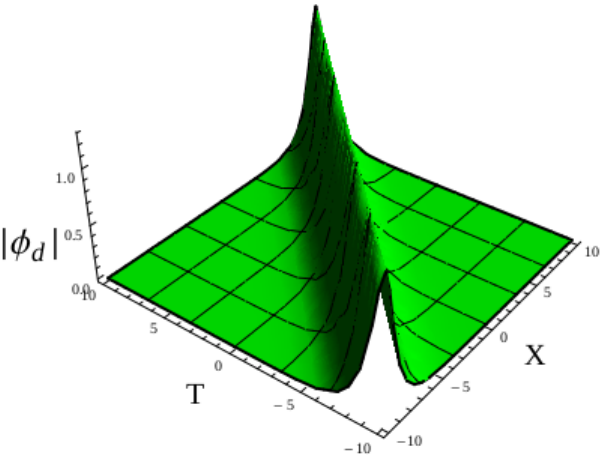}
    \caption{Quasi-deformed NLS single soliton.}
    \label{F1c}
\end{subfigure}\hspace{\fill} 
\begin{subfigure}[t]{0.5\textwidth}
    \includegraphics[width=\linewidth]{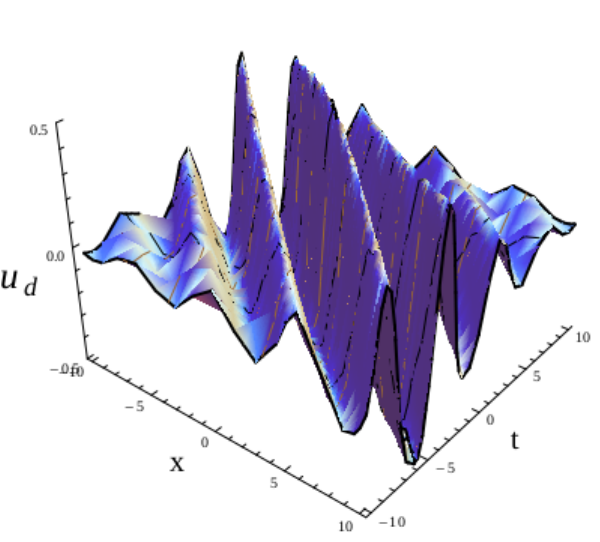}
    \caption{Quasi-deformed mapped KdV soliton train.}
    \label{F1d}
\end{subfigure}
\caption{The effect of quasi-deformation on the weak-coupling map from NLS to KdV solutions. The NLS single soliton (a) maps to a KdV soliton train (b), a property retained over the quasi-deformation though the effect of quasi-deformation is clear (c and d). Herein, $\epsilon=1.5$, $V=1=\rho$, $X_0=0=T_0$ $\varepsilon=0.1$ and $\omega_0=1=k_0$.}
\label{F1}
\end{figure}

It is not always possible to obtain solutions for arbitrary quasi-deformed systems like those in Eq.s \ref{N41}. However, since the QI parameter $\epsilon$ is independent, an order-by-order expansion can perturbatively lead to the quasi-deformed solution as seen for other systems \cite{FZ,FZ1,FZ2,N1,Own3}. The correspondence between the KdV and the NLS system at the solution level in a weak-coupling limit \cite{5} further supports this expectation. This correspondence materializes through the following parameterization of the NLS amplitude:

\bea
&&u=\varepsilon\left(\varphi e^{i\theta}+\bar{\varphi}e^{-i\theta}\right)+\frac{\varepsilon^2}{k_0^2}\left(\varphi^2 e^{i2\theta}+\bar{\varphi}^2e^{-i2\theta}\right)-2\frac{\varepsilon^2}{k_0^2}\vert\varphi\vert^2; \quad{\rm where},\label{31}\\
&&\theta=k_0x+\omega_0t, \quad 0<\varepsilon\ll 1, \quad \omega_0=k_0^3\neq 0.\nonumber
\eea
On substituting the above mapping in the KdV equation \ref{8}, equating terms with phase $e^{\pm i\theta}$ at ${\cal O}\left(\varepsilon^3\right)$, one arrives at the NLS equation,

\be
\varphi_T+i3k_0\varphi_{XX}+i\frac{6}{k_0}\vert\varphi\vert^2\varphi=0,\label{32}
\ee
and its complex conjugate with respect to the new coordinates,

\be
T=\varepsilon^2t \quad{\rm and}\quad X=\varepsilon\left(x+3k_0^2t\right).\label{34}
\ee
In Eq. \ref{32}, the `time'-derivative term comes from that of the KdV, the second derivative term comes from the third derivative term of the same and the non-linear term comes from its counterpart in KdV. Such direct correspondence, though approximate, strengthens the perturbative approach to obtain QI KdV system. One should keep in mind this analogical approach introduces {\it another} expansion parameter $\varepsilon$. This shows how the effect of quasi-deformation in one sector effects the other. The single bright soliton solution of a quasi-NLS system is of the form \cite{FZ},

\bea
&&\phi_d=\left[(2+\epsilon)^{1/2}\rho{\rm sech}\left\{(1+\epsilon)\rho\left(\tilde{X}-V\tilde{T}\right)\right\}\right]^{\frac{1}{1+\epsilon}}\exp\left[\left(\rho^2-\frac{V^2}{4}\right)\tilde{T}+\frac{V}{2}\tilde{X}\right],\nonumber\\
&&\qquad\tilde{X}=X-X_0,\quad\tilde{T}=T-T_0\quad{\rm and}\quad \rho,V,X_0,T_0\in{\mathbb R}_+\otimes{\mathbb R}\otimes{\mathbb R}\otimes{\mathbb R}.\label{FN001}
\eea
For the QID parameter $\epsilon\to 0$ one regains the undeformed soliton. Both these solutions are plotted in Fig.s \ref{F1a} and \ref{F1c} showing that the localized nature prevails over QID. The weak-coupling map of Eq. \ref{31} yields a soliton train-like solution for the KdV system (Fig. \ref{F1b}) that gets distorted over QID (Fig. \ref{F1d}) in the NLS sector. Tough it is not a priory guarantied that the weak-coupling map will persist over quasi-deformation it should be noted that the map itself is an approximation. Never the less, since the mapped KdV soliton train only displays minor local distortions over QID in the NLS sector, it can strongly be expected that quasi-KdV system can be obtained that supports localized solutions having a few conserved charges.

\paragraph*{}It should be more assuring to show that a proposed quasi-deformation of the KdV equation maps to a known quasi-NLS solution. For this purpose, we consider the case of Eq.s \ref{N41} with $m=3$ which essentially amounts the scaling of the nonlinear term as,  

\be
u_t+u_{xxx}+\left(6-\epsilon\right)uu_x=0,\label{35}
\ee
that maintains integrability as a special case of quasi-modification. This deformed KdV system invariably maps to a quasi-NLS system. From Eq.s \ref{31} the modified NLS equation is obtained as\footnote{In the map of Eq. \ref{31} \cite{5}, the non-linear terms of KdV and NLS systems map exclusively to each-other. Therefore, any scaling of the one in the KdV equation, like that in Eq. \ref{35}, corresponds to the same scaling of the similar term in the NLS equation.},

\be
\varphi_T+i3k_0\varphi_{XX}+i\left(1-\frac{\epsilon}{6}\right)\frac{6}{k_0}\vert\varphi\vert^2\varphi=0.\label{36}
\ee 
It is easy to see that,

\be
\left(1-\frac{\epsilon}{6}\right)\vert\varphi\vert^2\approx\frac{\delta}{\delta\vert\varphi\vert^2}V(\vert\varphi\vert) \quad{\rm with}\quad V(\vert\varphi\vert)\equiv\frac{1}{2}\vert\varphi\vert^{4\left(1-\tilde{\epsilon}\right)}, \quad\tilde{\epsilon}=\frac{\epsilon}{6},\label{37}
\ee
following the physical fact that the density $\vert\varphi\vert^2$ is sufficiently small in the weak-coupling limit. The above identification in terms of the NLS potential $V(\varphi)$ qualifies the obtained NLS system as a quasi-deformed one \cite{FZ2}. As long as the mapping prevails, the deformation of Eq. \ref{35} should represent a quasi-KdV system, which it trivially is, whose solutions can be obtained from those of the quasi-NLS ones as done above. More concretely this mapping between the two deformed systems justifies the Hamiltonian deformation approach to the quasi-KdV system. The appearance of essentially the same $sl(2)$ algebra for the quasi-NLS system \cite{FZ2}, though the linearizing sub-sector being different, further strengthens this line of argument.
\paragraph*{}Further confirmation of this assertion is obtained at the solution level. The single-soliton solution for the scaled KdV equation in Eq. \ref{35} has the form,

\be
u=\frac{c}{2}{\rm sech}^2\left[\frac{1}{2}\sqrt{c\beta}\left(x-x_0-\beta c(t-t_0)\right)\right],\quad{\rm where}\quad \beta=1-\frac{\epsilon}{3}\epsilon,\quad c>0,\label{Sol2}
\ee
which essentially amounts to a variable scaling of the undeformed system. The corresponding one-soliton solution for the NLS system of Eq. \ref{36} is,

\bea
&&\varphi(X,T)=K{\rm sech}\left[\Lambda_1K\left(\Lambda_2\tilde{X}-V\tilde{T}\right)\right]\exp\left[\frac{i}{2}\Lambda_2V\tilde{X}+\frac{i}{4}\left(\Lambda_1^2K^2-V^2\right)\tilde{T}\right];\label{Sol3}\\
&&{\rm where}\quad\Lambda_1=i\sqrt{\frac{3\beta}{k_0}},\quad\Lambda_2=-\frac{i}{\sqrt{3k_0}},\quad K\in\mathbb{R}_+,\nonumber
\eea
with similar variable scaling. These {\it exact} solutions could be obtained since both the deformed equations correspond to modification of the self-coupling strength maintaining integrability which may not be the case in general. However it strongly suggests that a quasi-KdV system can be obtained in the usual way that is known to work for other systems. Indeed the necessary algebraic framework obtained in the previous sections set up the conditions for a quasi-integrable KdV system. The anomaly ${\cal X}$ and deformed solution $u$ being parity-even should be enough for an {\it actual} quasi-KdV system.

\subsection{The perturbative expansion: An example}
\begin{figure}
\begin{subfigure}[t]{0.5\textwidth}
    \includegraphics[width=\textwidth]{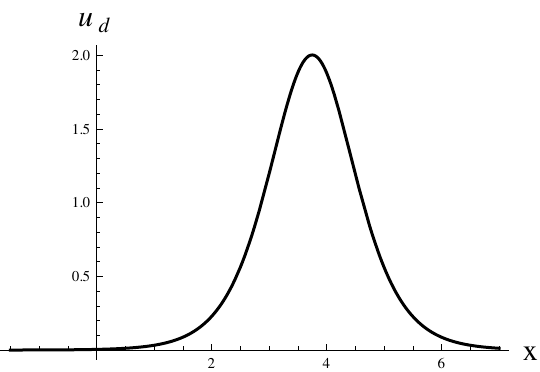}
    \caption{$\epsilon$=0}
\label{F2a}
\end{subfigure}\hspace{\fill} 
\begin{subfigure}[t]{0.5\textwidth}
    \includegraphics[width=\linewidth]{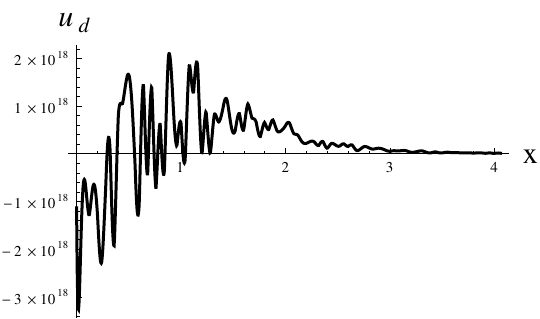}
    \caption{$3\epsilon=0.1$.}
\label{F2b}
\end{subfigure}
\begin{subfigure}[t]{0.5\textwidth}
    \includegraphics[width=\linewidth]{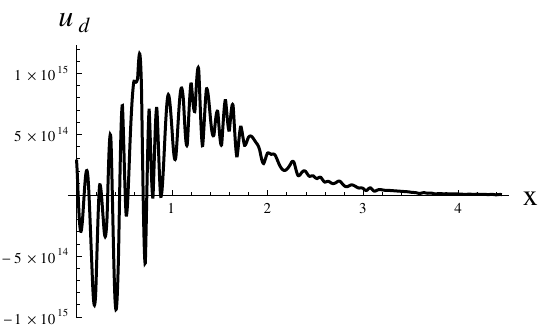}
    \caption{$3\epsilon=0.2$.}
    \label{F2c}
\end{subfigure}\hspace{\fill} 
\begin{subfigure}[t]{0.5\textwidth}
    \includegraphics[width=\linewidth]{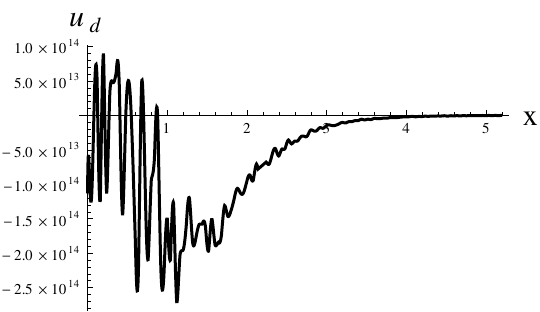}
    \caption{$3\epsilon=0.5$.}
    \label{F2d}
\end{subfigure}
\caption{Numerical solutions $u_d$ corresponding to the deformation $u^3\rightarrow u^{3++3\epsilon}$ in the Hamiltonian. The very localized and parity-even single-soliton structure for $\epsilon=0$ (\ref{F2a}) gets significantly distorted (\ref{F2b}, \ref{F2c}, \ref{F2d}) even for small values $\epsilon$. They still remain fairly localized suggesting asymptotic conservation of the corresponding charges. These plots are evaluated at time $t=1$.}
\label{F2}
\end{figure}

\begin{figure}
\centering
\includegraphics[width=0.6\textwidth]{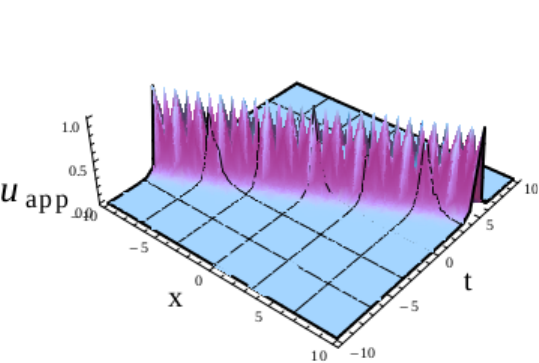}
\caption{Approximate soliton train-like solution for power-scaling $u\rightarrow u^{1+\epsilon}$ of the nonlinear term in the Hamiltonian with $\epsilon=0.7$ and $c_0=1$.}
\label{F3}
\end{figure}

Choosing different forms of $F(u)$ to obtain a parity-even ${\cal X}$ ensures asymptotic (or quasi) integrability, which also includes some integrable systems. For example a choice of $F(u)=\kappa_3u^3+\kappa_4u^4$ results in a parity-even ${\cal X}$. However, it leads to Gardner or mKdV equations depending on the values of $\kappa_{3,4}$ which are completely integrable and further supports kink-type solutions. As a non-trivial example, instead of an extension to the Hamiltonian like $F(u)$, we consider a power-modification of the nonlinear term therein by the QID parameter $\epsilon$ in the same spirit of the quasi-NLS system \cite{FZ2} as,

\be
H^{\rm Def}_1[u]\equiv\int_{-\infty}^\infty dx~\left(\frac{1}{2}u_x^2-u^{3+3\epsilon}\right).\label{N22}
\ee
It amounts to deforming the KdV nonlinearity through power-scaling of the amplitude $u^3\rightarrow u^{3+3\epsilon}$. As the nonlinearity has directly been effected the corresponding equation,

\be
u_t+u_{xxx}+6uu_x=4uu_x-2(1+\epsilon)(2+3\epsilon)u^{1+3\epsilon}u_x,\label{A1}
\ee
becomes relatively difficult to solve. We have numerically obtained a few solutions using {\it Mathematica 8} for different values of $\epsilon$ in Fig.s \ref{F2} which represent deviations from the undeformed structure. As expected for finite $\epsilon$ the solutions do not posses definite parity which eventually distorts the parity of the anomaly function ${\cal X}$. This results in non-conserved charges at any finite time. However, these deformed solutions are still localized, which could mean strong interactions or radiation-effected solitonic structures \cite{KdVQI1}, strongly suggesting asymptotic conservation of the same charges.
To have a better idea about this system we undertake an order-by-order expansion of this system in terms of $\epsilon$. Though this parameter need not be small always, such an order-expansion is valid for the dependence of the solution on the parameter $\epsilon$ being {\it analytic} \cite{FZ2}. For the particular case, such an expansion of the anomaly takes the form,

\be
{\cal X}=2u^2-2(1+\epsilon)u^{2+3\epsilon}\approx-2\epsilon u^2\left(1+3\ln(u)\right)+{\cal O}(\epsilon^2),\label{N23}
\ee
which mirrors the parity of the solution. It provides a logarithmic nonlinearity at first order in $\epsilon$ making the evaluation of quasi-corrections at that order quite difficult, especially when $u\ll1$. As an approximation, for finite $u$ and $\epsilon\ll1$, a localized solution has the form $u_{app}=\frac{c_0}{2}{\rm sech}^2\phi$ where $\phi$ satisfies the following approximate expression:

\be
x-c_0t\approx-\frac{2}{\sqrt{c_0}}(1-2\epsilon)\phi+\frac{2}{\sqrt{c_0}}\epsilon\left(\ln\frac{c_0}{2}+2\ln{\rm sech}\phi\right)\coth\phi.\label{2N08}
\ee
Clearly, it goes to the usual KdV bright soliton for $\epsilon=0$. A plot for the deformed solution $u_{app}$ in Fig. \ref{F3} depicts a soliton train-like structure. Such structures strongly suggest asymptotic integrability of the system if not actual integrability. 

\begin{figure}
\centering
\includegraphics[width=0.6\textwidth]{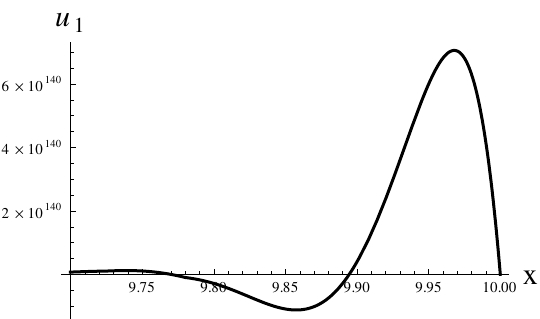}
\caption{${\cal O}(\epsilon)$ correction to the deformed KdV solution. It clearly deviates from even-parity structure eventually leading to non-conserved charges.}
\label{F4}
\end{figure}

In order to evaluate the charges and analyze their (quasi-)conservation we consider an order-by-order expansion in $\epsilon$ of all the quantities of the deformed system. We assume quantities like the solution $u$, anomaly ${\cal X}$, charge $Q^n$ etc to be fairly well-behaved functions of $\epsilon$ and thus can be expanded as power-series in the same. The deformed solution can thus be expanded as,

\be
u=u_0+\epsilon u_1+\epsilon^2u_2+\cdots,\label{2N09}
\ee
with $u_0$ satisfying the undeformed KdV equation. Similarly the anomaly and rate of change of charges can be expanded as,

\be
{\cal X}=\epsilon{\cal X}_1+\epsilon^2{\cal X}_2+\cdots\quad{\rm and}\quad \Gamma^n=\epsilon \Gamma^n_1+\epsilon^2 \Gamma^n_2+\cdots
\ee
The zero-order contribution to the anomaly vanishes as it corresponds to the undeformed system and so does that to $\Gamma^n$. For the anomaly in Eq. \ref{N23} the ${\cal O}(\epsilon)$ contribution to the deformed equation has the form:

\be
u_{1,t}+u_{1,xxx}+6\left(u_0u_1\right)_x=-(10+12\ln u_0)u_0u_{0,x},\label{2N10}
\ee
The solution $u_1$ of this equation, with $u_0$ being the KdV 1-soliton solution, is depicted in Fig. \ref{F4}. It does not have definite parity unlike the parity-even undeformed solution in Fig. \ref{F2a}, thus yielding a first-order deformed solution $u_d=u_0+\epsilon u_1$ without definite parity. Consequently, the anomaly for this particular case can be expanded as,

\be
{\cal X}=-2\epsilon\left\{1+3\ln(u_0)\right\}u_0^2+3\epsilon^2\left\{u_0u_1+(2u_1+u_0)u_0\ln(u_0)\right\}+{\cal O}\left(\epsilon^3\right),\label{A2}
\ee
which is no longer parity-even at ${\cal O}\left(\epsilon^2\right)$. 
As for the rate of change of the charges $\Gamma^n$ the ${\cal O}(\epsilon)$ depends only on $u_0$ after expanding the coefficients $f_n^+$s in $\epsilon$ and thus vanishes. One can trivially check this for the KdV single soliton $u_0=(c_0/2){\rm sech}^2\left[\sqrt{c_0}\left(x-c_0t\right)/2\right]$ as $f_n^+$ contain only odd derivatives of $u$ (Eq.s \ref{N010}). As a demonstration, the next order contribution for $n=2$ is,

\be
\Gamma^2_2=\frac{1}{2\sqrt{2}}\int_x\left[{\cal X}_1u_{1,x}+{\cal X}_2u_{0,x}\right]\quad{\rm since}\quad f_{-2}^+=\frac{1}{2\sqrt{2}}\left(u_0+\epsilon u_1\right)_x,\label{A3}
\ee
where ${\cal X}_1$ and ${\cal X}_2$ can be read off of Eq. \ref{A2}. Clearly, $\Gamma^2_2\neq0$ since it contains $u_1$and thus $Q^{-2}$ will not be conserved eventually.
\paragraph*{}In principle these deviations from integrability can be calculated exactly for all $n$. The ${\cal O}(\epsilon)$ contributions to ${\cal X}$ and the integrand in $\Gamma^n$ are entirely constituted of the undeformed solution $u_0$ whereas the ${\cal O}\left(\epsilon^2\right)$ contributions contain only $u_1$ in addition. The corrections to the undeformed solution $u_n$s can successively be calculated from the ${\cal O}\left(\epsilon^n\right)$ contribution to the parent equation \ref{A1} after evaluating $u_{n-1}$ previously and thus, eventually, all the deformed charges and their rates can be obtained up to all orders in principle. A good convergence of the net sum of these contributions should ensure quasi-conservation but it requires a good deal of numeric simulation which is beyond this work. However, such confirmation has already been obtained for particular quasi-KdV systems \cite{KdVQI1}.

\subsection{Connection with Non-Holonomic Deformation}
It is fruitful to compare the implications of quasi-deformation obtained thus far with those from {\it nonholonomic deformation} of the KdV and as well as that of the NLS systems. The nonholonomic deformation
is practically obtained through extending the temporal Lax component with local functions of various grade by infusing particular powers of the spectral parameter $\lambda$ into them, which does not effect the time-evolution of the system. This incorporates an inhomogeneous extension to the original differential equation, with higher order differential constraints imposed on the deformation functions, obtained through retaining the zero-curvature condition. Thus the deformed system still stays
{\it integrable}. Nonholonomic deformation had been well-analyzed for KdV and coupled complex KdV systems \cite{1}, from both loop-algebraic  
and AKNS approaches, and in case of NLS systems it has recently been shown that the nonholonomic deformation is locally different from quasi deformation as the latter leads to non-integrable, but asymptotically they may converge \cite{AGM}. 

\paragraph*{}The quasi-deformation is usually applied at the 
level of functions of the independent variable \cite{FZ,FZ1,FZ2,N1} or that of a functional as in the present case for KdV, that deforms the 
Lax component itself without effecting at the other spectral sectors. This generally yields a non-zero curvature. Therefore, both the deformations are fundamentally different. However, since the quasi-conserved charges asymptotically are conserved a quasi-deformed system may converge to a nonholonomic one asymptotically. In the present case the deformed solutions maintain localization. Further, in the present case, the logarithmic contribution ({\it e. g.} in Eq. \ref{N23}) becomes subdominant at large distances for a localized deformed solution $u$ leaving a pure KdV-type system with scaled constants which should be integrable. A similar property was observed for the NLS system in Ref. \cite{AGM}. Considering the weak coupling map between these two systems and their quasi-deformations (Sec. \ref{Sec4}) one could expect the quasi-KdV system to converge to a nonholonomic variant of the KdV system. 

\paragraph*{}In that regime it may be possible to interpret the present power-series expansion in $\epsilon$ as local constraints characterizing a nonholonomic system, that are identified with constraints. The present quasi-KdV system further supports single- and multi-soliton-type solutions which can very well converge to their ideal counterparts asymptotically. Thus it will be interesting to identify such systems with order-by-order relations (`constraints') by evaluating asymptotic form of the exact solution for quasi-KdV and other systems.

\section{Conclusions and Discussion}\label{Sec5}
It is seen that a comprehensive quasi-integrable deformation of the KdV system is indeed possible, provided the loop-algebraic generalization \cite{1} has been considered. As the KdV equation is not dynamical neither in the sense of Galilean (like NLS) nor Lorentz (like SG) systems, the deformation has to be performed at an {\it off-shell} level ({\it i. e.}, without using the EOM). The available Hamiltonian formulation of KdV system comes to rescue in this {\it ab-initio} treatment, wherein the Lax construction in the $SU(2)$ representation has been utilized to obtain the standard Abelianization that utilized the inherent $sl(2)$ loop algebra. In the sub-space where the Abelianization manifests, anomalous conservation laws were obtained. The anomaly function and the coefficients of the rotated Lax components are shown to have definite parity properties given the deformed solution being a parity eigenstate, subsequently ensuring asymptotic conservation of the anomalous charges implying quasi-integrability.
\paragraph*{}As particular cases, both local extensions as well as power-deformation in the Hamiltonian density are considered. The prior allows for constructing a scaled KdV at the simplest level, with single-soliton profile, as well as families of higher-derivative extensions to the same with some of them possibly being quasi-integrable, with at least one conserved charge. This is intuitively allowed, following the weak-coupling correspondence
between KdV and NLS systems. The compatibility of the present deformation with that of QI NLS system has been obtained, followed by corresponding one-soliton solutions with variable scaling. The deformed KdV solution corresponding to the quasi-NLS soliton appeared to have similar properties to the KdV soliton train that corresponds to the undeformed NLS soliton.
In case of power deformation of the type $u\to u^{1+\epsilon}$ in the Hamiltonian the situation becomes more complicated with possible singularities. Single soliton-like localized structures are still supported by these deformed systems. Further, an order-by-order expansion in the deformation parameter led to localized solutions. Although they may not posses definite parity, asymptotically they are expected to yield conservation of the charges. It will be worthwhile to numerically analyze particular stable solutions to this deformed KdV and higher derivative systems, and to study their behavior with those from QI NLS system when the weak correspondence is valid. We aspire to imply the same for complex coupled KdV formalism in the future.
\paragraph*{}The present approach of quasi-deformation can be extended to further related systems, including KdV-type hierarchies, mKdV and their non-local counterparts. An obvious generalization would be that of the coupled complex KdV system of Eq.s \ref{25} could be more challenging owing to the requirement of a constant semi-simple $sl(2)$ element in the spatial Lax component. A suitable form of deformation of this component could be,

\be
A\to\left(\frac{\bar{q}}{\lambda}+1\right)\sigma_+-(u-\lambda)\sigma_-\equiv\frac{\bar{q}-q}{\sqrt{2}}F_+^{-1}+\frac{\bar{q}+q}{\sqrt{2}}F_-^{-1}+\sqrt{2}F_+^0,\label{2N14}
\ee
under the $sl(2)$ representation of Eq.s \ref{20}. However, this includes a scaling of the conjugate field $\bar{q}$ by the spectral parameter which violates its relative grading with respect to $q$ in $A$ to yield a KdV system, the latter being apparent from the following equivalent Lax pair:

\bea
&&A\to\frac{\bar{q}}{\lambda}\sigma_+-\lambda q\sigma_-\quad{\rm and}\nonumber\\
&&B\to\left(\bar{q}q_x-q\bar{q}_x\right)\sigma_3-\frac{1}{\lambda}\left(\bar{q}_{xx}+2q\bar{q}^2\right)\sigma_++\lambda\left(u_{xx}+2\bar{q}q^2\right)\sigma_-,\label{2N15}
\eea
with explicit spectral dependence, yielding the same undeformed KdV system. This hampers the Abelianization scheme, more so as the Lax pair of Eq.s \ref{23} (or Eq. \ref{2N15}) is a direct consequence of the AKS hierarchy \cite{1}. However the quasi-deformation of this complex coupled KdV system may be possible through some non-trivial Lax representation; which may directly be obtained through brute-force numerical calculations.
\paragraph*{}The aim of the present work was to obtain a first-principle quasi-deformation formalism of the KdV system which corresponds to deformation of the corresponding Hamiltonian. This has led to various possible deformed structures which may be quasi-integrable. The particular case of RLW and mRLW systems are known to be so \cite{KdVQI1} and multi-solitonic structures were numerically obtained in conformity. However, pure analytic determination ({\it e. g.} by Hirota method) of definite-parity localized solutions of these systems are yet to be achieved. The perturbative approach of the present work supply some insight into the possible solutions and their localization is found to be very much possible. A full-on numerical simulation of such solutions is beyond the scope of this work. We expect to take up this task in the near future. We further aspire to analyze similar systems like mKdV and other hierarchies of the KdV system for possible quasi-deformation and to extend this study to non-local systems.

\vskip 0.5cm
\noindent{\it Acknowledgement:} Kumar Abhinav's research is supported by Mahidol University,
Thailand under the Grant Number MRC-MGR 04/2565. Partha Guha is grateful to Jun Nian and Vasily Pestun for interesting discussions. The authors are also grateful to Professors Luiz. A. Ferreira, Wojtek J. Zakrzewski and Betti Hartmann for their encouragement, various useful discussions and critical reading of the draft during the initial phases of this work.

\end{document}